\documentstyle[12pt,aaspp4,flushrt]{article}
\input{psfig}

\makeatletter

\@ifundefined{chapter}{\def\thebibliography#1{\section*{References\@mkboth
  {REFERENCES}{REFERENCES}}\list
  {\relax}{\setlength{\labelsep}{0em}
        \setlength{\itemindent}{-\bibhang}
        \setlength{\itemsep}{0pt}
        \setlength{\parsep}{0pt}
        \setlength{\leftmargin}{\bibhang}}
    \def\newblock{\hskip .11em plus .33em minus .07em}
    \sloppy\clubpenalty4000\widowpenalty4000
    \sfcode`\.=1000\relax}}%
{\def\thebibliography#1{\chapter*{Bibliography\@mkboth
  {BIBLIOGRAPHY}{BIBLIOGRAPHY}}\list
  {\relax}{\setlength{\labelsep}{0em}
        \setlength{\itemindent}{-\bibhang}
        \setlength{\itemsep}{0pt}
        \setlength{\parsep}{0pt}
        \setlength{\leftmargin}{\bibhang}}
    \def\newblock{\hskip .11em plus .33em minus .07em}
    \sloppy\clubpenalty4000\widowpenalty4000
    \sfcode`\.=1000\relax}}

\newlength{\bibhang}
\let\@internalcite\cite
\def\cite{\let\@citeleft(\let\@citeright)%
    \@ifstar{\citeyear}{\citefull}}
\def\citenp{\let\@citeleft\relax\let\@citeright\relax
    \@ifstar{\citeyear}{\citefull}}
\def\citefull{\def\astroncite##1##2{##1~##2}\@internalcite}
\def\citeyear{\def\astroncite##1##2{##2}\@internalcite}

\def\@citex[#1]#2{\if@filesw\immediate\write\@auxout{\string\citation{#2}}\fi
  \def\@citea{}\@cite{\@for\@citeb:=#2\do
    {\@citea\def\@citea{; }\@ifundefined
       {b@\@citeb}{{\bf ?}\@warning
       {Citation `\@citeb' on page \thepage \space undefined}}%
{\csname b@\@citeb\endcsname}}}{#1}}

\def\@cite#1#2{\@citeleft#1\if@tempswa , #2\fi\@citeright}
\def\@biblabel#1{}

\makeatother

\newcommand\approxlt{\mbox{$^{<}\hspace{-0.24cm}_{\sim}$}}

\def\bi#1{\hbox{\boldmath{$#1$}}}

\begin{document}

\title{Weighing the Cosmological Energy Contents\\ with Weak
Gravitational Lensing} 
\author{Lam Hui\altaffilmark{1}}
\affil{NASA/Fermilab Astrophysics Center\\ Fermi
National Accelerator Laboratory, Batavia, IL 60510}

\altaffiltext{1}{e-mail: \it lhui@fnal.gov}

\begin{abstract}
Bernardeau et al. (1997), using perturbation theory, showed that the
skewness of the large-scale lensing-convergence, or projected mass
density, could be used to constrain $\Omega_m$, the matter content of
the universe. On the other hand, deep weak-lensing field surveys in the near
future will likely measure the convergence on small angular scales
($\, \approxlt \, 10$ arcmin.), where the signal will be dominated by 
highly nonlinear fluctuations. We develop a method to compute the
small-scale convergence skewness, making use of a prescription for 
the highly nonlinear three-point function developed by Scoccimarro and
Frieman (1998). This method gives predictions that agree well with 
existing results from ray-tracing N-body simulations, but is 
significantly faster, allowing the exploration of a large number
of models. We demonstrate that the small-scale convergence
skewness is insensitive to the shape and normalization of
the primordial (CDM-type) power spectrum, making it dependent almost entirely
on the cosmological energy contents, through their influence on
the global geometrical distances and fluctuation growth rate. 
Moreover, nonlinear clustering appears to enhance the differences
between predictions of the convergence skewness for a range of
models. Hence, in addition to constraining $\Omega_m$, the small-scale
convergence skewness from future deep several-degree-wide surveys can
be used to differentiate between curvature 
dominated and cosmological constant ($\Lambda$) dominated models, as well as 
to constrain the equation of state of a quintessence component, 
thereby distinguishing $\Lambda$ from quintessence as well. 
Finally, our method can be easily generalized to other measures 
such as aperture mass statistics.
\end{abstract}

\keywords{cosmology: theory
--- gravitational lensing --- large-scale structure of universe}

\section{Introduction}
\label{intro}

The correlated shear of images of distant galaxies provides a promising
way to probe the intervening large scale structure of the universe
(e.g. \citenp{blandford91,miralda91,kaiser92};
see also \citenp{jsw99} [JSW hereafter] and ref. therein).
The convergence can be constructed from a shear map (\citenp{ks93}), which
can be interpreted as a form of projected mass density (\citenp{kaiser92}; our
notation follows that of \citenp{js97}):
\begin{equation}
\kappa({\bi \theta})=
\int_0^{\chi_s} d\chi\ w(\chi)\ \delta(r(\chi){\bi \theta},\chi)
\label{kappadef}
\end{equation}
where $\delta$ is the mass overdensity as a function of spatial position
(and implicitly as a function of time as well, where the time
and space coordinates fall on the photon null geodesic), ${\bi
\theta}$ is the angular position on the sky,  
$\chi$ is the comoving distance along the line of sight, $r(\chi)$ is
the angular diameter distance, and
$w(\chi)$ is a weight function which depends on a combination
of global geometrical distances and is proportional to the total
matter density of the universe. The coordinates
$\chi = 0\, , \, \chi_s$ denote respectively the positions of the observer
and the sources or background galaxies. The so-called Born approximation has
been assumed
(see \citenp{bern97} for a discussion). 

It is clear that $\kappa$ is a valuable quantity for
cosmology. One can derive important constraints on the
cosmological density parameters 
$\Omega$'s through the dependence of $\kappa$ on the global
geometrical distances and the evolution of $\delta$ on the line cone.
A commonly used statistic is its second moment
$\langle \kappa^2 \rangle$, or the two-point correlation function
$\langle \kappa ({\bi \theta}) \kappa ({\bi \theta'}) \rangle$. However, it is
clear that the second moment depends on the mass power spectrum as
well (e.g. \citenp{js97,kaiser98}). 

Bernardeau et al. \cite*{bern97},
using perturbation theory, showed that this degeneracy could be broken by using
the convergence skewness $S_3 \equiv \langle \kappa^3 \rangle /
\langle \kappa^2 \rangle^2$. 
It is customary to consider $S_3$ as a
function of angular scale $\theta_R$, assuming $\kappa$ is first
smoothed on scale $\theta_R$. However, future weak lensing surveys are
likely to yield measurements 
of $S_3$ first on small angular scales, $\theta_R < 10'$, both
because the small-scale shear signal is stronger and also because
of larger sampling fluctuations on large scales (see e.g.
\citenp{ludo99}, JSW) . For 
sources at a redshift of $z = 1$, the 
peak contributions to the lensing signal will
come from $z \sim 0.5$, which for $\theta_R <
10'$ translates into a comoving length scale of less than a few Mpc
that is generally 
comparable to or smaller than the nonlinear scale (where the rms
density fluctuation is of order unity). It
is therefore expected 
that the perturbative treatment of Bernardeau et al. \cite*{bern97}
would not hold for these angular scales of interest.
This has in fact been explicitly demonstrated by JSW
using the technique of ray-tracing N-body simulations (see their Fig.
18; see also \citenp{couchman98}). Unfortunately, the prediction of
skewness from N-body
simulations can become prohibitively
expensive, if one is interested in exploring a
large number of cosmological models.

There is therefore a need for alternative methods to predict
accurately and efficiently the small-scale skewness. It is our aim
here to develop 
such a method. We will test 
it by comparing with existing results from simulations, and
show that the small-scale skewness is a sensitive probe of $\Omega$'s.
We will then apply it 
to cosmological models that have not been considered before in the
context of weak lensing. In particular, we will predict the small
angular-scale skewness for a quintessence model, where quintessence
is a component of the cosmological fluid that has negative pressure
(e.g. \citenp{pr88,frieman95,coble97,tw97,fj98,caldwell98}).
Such models, which include the cosmological constant dominated models
as a limiting case, are currently in favor in part because
of recent Type Ia supernova measurements
(\citenp{riess98,garnavich98,perlmutter98}). We will demonstrate that
the convergence skewness can provide interesting constraints on them.

\section{The Convergence Skewness}
\label{skewness}

Let us first give the
expressions for the cosmology dependent geometrical quantities that 
appear in eq. (\ref{kappadef}). 
The comoving 
distance along the line of sight $\chi$ is given by
(\citenp{peebles93})
\begin{equation}
\chi(z) = c H_0^{-1} \int_0^z {dz' [\Omega_m (1+z')^3 +
\Omega_k (1+z')^2 + \Omega_q (1+z')^{3(1+w_q)}]^{-1/2}}
\label{chidef}
\end{equation}
where $z$ is the redshift of interest, $H_0$ is the Hubble parameter
today, $c$ is the speed of light, and the $\Omega$'s denote the
fractions of the critical energy 
density today in various components: $\Omega_m$ for pressureless matter or
dust, $\Omega_k$ for spatial curvature and $\Omega_q$ for quintessence
or a fluid with negative pressure (its pressure $p$ is related to
its density $\rho$ by $p = w_q \rho$, where $w_q < 0$), with the
cosmological constant $\Lambda$ as a limiting case ($w_q = -1$). The
$\Omega$'s sum to unity.
The angular-diameter distance $r(\chi)$ is given by
$r(\chi)=K^{-1/2}\sin K^{1/2}\chi, \chi, (-K)^{-1/2}\sinh
(-K)^{1/2}\chi$ for closed, flat and open models respectively, and
$K=(\Omega_m-1)c^{-2} H_0^2$.  
In other words, the 
metric is given by $ds^2 = -c^2 dt^2 + a(t)^2 (d\chi^2 + r(\chi)^2 d^2
\theta)$, where $a(t) = 1/(1+z)$ is the expansion scale factor as a function of
proper time $t$. The line-of-sight projection of $\delta$ in eq.
(\ref{kappadef}) is weighed by the
function $w(\chi)$: 
\begin{eqnarray}
\label{wchidef}
w(\chi)&=&{3\over 4a} c^{-2} H_0^2 \Omega_m 
{r(\chi)\ r(\chi_s - \chi) 
\over r(\chi_s)}
\end{eqnarray}
where $\chi_s$ is the comoving radial position of the sources.
Note that here, as in the rest of the paper, we assume
all sources are at the same redshift. Eq. (\ref{kappadef}) and
(\ref{wchidef}) 
can be easily generalized to the case of multiple source-redshifts 
by integrating over contributions from different $\chi_s$'s. 
Statistical measures of $\kappa$ for sources distributed in a
realistic fashion can usually be approximated by having all
sources at the same mean redshift (e.g. JSW).

The convergence skewness is defined by
\begin{equation}
S_3 (\theta_R) \equiv {\langle {\kappa_{\theta_R}}^3 \rangle \over \langle
{\kappa_{\theta_R}}^2 \rangle^2 } \, \, , \, \, \kappa_{\theta_R}
\equiv \int \kappa({\bi \theta'}) W_{\theta_R} ({\bi \theta}
- {\bi \theta'})
d^2 \theta'
\label{S3def}
\end{equation}
where $W_{\theta_R}$ is a smoothing kernel of radius $\theta_R$ (in
this paper, we will use a top-hat).
The utility of $S_3$ derives from, crudely speaking, the fact that its
analogue for the mass overdensity ($\langle \delta^3 \rangle / \langle \delta^2
\rangle$) is quite insensitive to details of the power spectrum,
especially on small scales.
Hence, as we will see, $S_3$ is almost purely determined by the
cosmological energy contents.

Combining eq. (\ref{kappadef}) and (\ref{S3def}), it can be shown that
\begin{eqnarray}
\label{S3cal}
&& S_3 = K_3 / (K_2)^2 \quad \quad \quad {\rm with}\\ \nonumber
&& {K}_3 \equiv (2\pi)^2 \int_0^{\chi_s} d\chi {w(\chi)^3 \over r(\chi)^4}
\int {d^2 \ell_1} {d^2 \ell_2}
B({\bi \ell_1}/r(\chi), {\bi \ell_2}/r(\chi), {\bi \ell_3}/r(\chi)) 
\\ \nonumber && \quad \quad \quad \tilde W (\ell_1 \theta_R) \tilde W
(\ell_2 \theta_R) \tilde W (\ell_3 \theta_R) \\ \nonumber
&& {K}_2 \equiv 2\pi \int_0^{\chi_s} d\chi {w(\chi)^2 \over r(\chi)^2}
\int {d^2 \ell} P(\ell/r(\chi)) \tilde W (\ell
\theta_R)^2
\end{eqnarray}
where $\tilde W (x)$ is the Fourier transform of the two-dimensional
top-hat, $\tilde W (x) = 2 J_1 (x) /x$ with $J_1$ being the
first-order Bessel function.
The $\ell$'s represent
the Fourier coordinates in angular space, in other words we
are taking the small-angle approximation where spherical harmonics
can be replaced by plane waves. The combination ${\bi \ell_1
+ \bi \ell_2 + \bi \ell_3}$ vanishes. The three-dimensional mass power
spectrum and 
bispectrum are respectively $P$ and $B$. Our convention is:
$\xi_2 (|{\bf r}|) = \int d^3 k P(k) e^{-i{\bf k}\cdot{\bf r}}$ and
$\xi_3 ({\bf r_1}, |{\bf r_2}|)
= \int d^3 k_1 d^3 k_2 B({\bf k_1}, {\bf k_2}, - {\bf k_1} - {\bf k_2}) 
e^{-i {\bf k_1}\cdot{\bf r_1} - i {\bf
k_2}\cdot{\bf r_2}}$ where the ${\bf k}$'s denote the Fourier coordinates
in three-dimensional space, and $\xi_2$ and $\xi_3$ are the two-
and three-point correlation functions respectively. 
(For readers who
are used to putting $(2\pi)^3$ under $d^3 k$: simply replace
all relevant expressions in this paper by $P \rightarrow P/(2\pi)^3$ and
$B \rightarrow B/(2\pi)^6$.)
Note how in eq.
(\ref{S3cal}) the projection forces the ${\bf k}$'s to lie in the
plane of the sky.
The reader is referred to Kaiser \cite*{kaiser92}, Bernardeau et al.
\cite*{bern97} and Jain \& Seljak \cite*{js97} for detailed
derivations. 

To compute $S_3$ on small angular scales, we need
to understand the nonlinear evolution of $P$ and $B$. 
The nonlinear behavior of $P$ can be described by a scaling
ansatz introduced by Hamilton et al. \cite*{hamilton91}, which was
later extended by Jain et al. \cite*{jain95} and Peacock \& Dodds
\cite*{pd94,pd96}. We will employ the latest version set
out in the 
latter. Jain \& Seljak \cite*{js97} have considered the two-point
version of $K_2$ using this ansatz. Essentially, the ansatz consists
of postulating that 
$ 4 \pi k^3 P(k) = f [ 4 \pi k_L^3 P_L (k_L) ]$ where
$f$ is some universal function, and
$P_L (k_L)$ is the linear power spectrum at the rescaled wave-number
defined by $k_L = [1+ 4 \pi k^3 P(k) ]^{-1/3} k$. 
The cosmological dependence comes in through the linear fluctuation
growth rate $P_L \propto [g(z)/(1+z)]^2$ (fitting formula from
\citenp{carroll92}): 
\begin{equation}
g(z) = {5\over 2} \Omega_m (z) [\Omega_m (z)^{4/7} - \Omega_\Lambda (z) +
(1+\Omega_m (z)/2) (1+\Omega_\Lambda (z)/70)]^{-1}
\label{gdef}
\end{equation}
where $\Omega_m (z)= \Omega_m (1+z)^3 / [\Omega_m (1+z)^3 + \Omega_k
(1+z)^2  +
\Omega_\Lambda]$ and $\Omega_\Lambda (z) = 
\Omega_\Lambda / [\Omega_m (1+z)^3 + \Omega_k (1+z)^2 + \Omega_\Lambda]$,
and $\Omega$'s without explicit $z$ dependence denote their values
today. For quintessence models with $w_q \ne -1$, we integrate numerically the
equation for the linear growth rate, and substitute this
in the corresponding expressions given by Peacock \& Dodds
\cite*{pd96} (see \citenp{wang98} for a useful fitting formula).

For the bispectrum, it has been conjectured for some time that
the following scaling approximately holds in the highly nonlinear regime (e.g. 
\citenp{DP77,peebles80,fry84,hamilton88}):
\begin{equation}
B({\bf k_1}, {\bf k_2}, {\bf k_3}) = Q_3 (P(k_1) P(k_2) + P(k_2) P(k_3)
+ P(k_3) P(k_1))
\label{Q3def}
\end{equation}
where the three ${\bf k}$'s form a closed triangle, and
$Q_3$ is a weak function of scale but independent of the triangle
configuration. The above nonlinear hierarchical form (sometimes called
the hierarchical ansatz) is also observed in N-body
simulations (see \citenp{roman98} and ref. therein).
The problem, however, was that, there has been for a long time no way
to predict the amplitude of $Q_3$, other than by examining N-body
simulations on a case by case basis. Recently, Scoccimarro \& Frieman
\cite*{sf98} introduced a method they named hyperextended perturbation
theory which allows one to calculate $Q_3$ analytically:
\begin{equation}
Q_3 (n) = [4 - 2^n]/[1+2^{n+1}]
\label{Q3}
\end{equation}
where $n$ is the {\it linear} power spectral index at the scale of
interest $(k_1 + k_2 + k_3)/3$. The above expression implies that
$Q_3$ is insensitive to cosmology, except through $n$ (see 
\citenp{roman98}). 

Hence, combining eq. (\ref{Q3def}), (\ref{Q3}) and the nonlinear
evolution of $P$ given by Peacock \& Dodds \cite*{pd96}, together with
eq. (\ref{S3cal}), completely specifies $S_3$ for any given
primordial power spectrum and cosmology. To ease the computation,
we find that the following approximation for $K_3$ agrees with the
exact integration to within a few percent for the models considered in
this paper:
\begin{equation}
K_3 \sim 3 (2\pi)^2 \int_0^{\chi_s} d\chi {w(\chi)^3\over r(\chi)^4}
\left[ \int d^2 \ell \sqrt Q_3 P(\ell/r(\chi)) \tilde W (\ell
\theta_R)^2 \right]^2
\label{K3approx}
\end{equation}
where $Q_3$ is evaluated at an $n$ corresponding to the scale $\ell/r(\chi)$. 
This approximation works in part because $Q_3$ varies slowly with scale,
on the relevant small scales.

In Fig. \ref{skew_comp}a, we show a comparison of the skewness
computed as described above with the skewness obtained from
ray-tracing N-body simulations (JSW), for sources at $z = 1$. The error-bars
shown are estimated from the dispersions between 5-10 ray-tracing
realizations. Three models
are shown (see Table \ref{table}). They are all normalized to
match the cluster abundance today.
The agreement is good, to better than $10 \%$ for $\theta_R \sim
1' - 5'$, and it remains reasonable at larger angular scales, although
its exact level is somewhat uncertain because of the large dispersions
of the N-body results. 
The agreement here is to be contrasted with the as much as $30 \%$ 
discrepancy for the perturbation theory predictions, shown
as points on the far left of Fig. \ref{skew_comp}a. In particular,
perturbation theory brings $S_3$ for OCDM and LCDM much closer
than what it should be (the good agreement of the perturbative $S_3$
with the actual value for LCDM seems to be a coincidence). It is
interesting how nonlinear clustering 
makes it easier to tell them apart. There seems to be
a complicated interplay of projection and nonlinear clustering (e.g. 
\citenp{gb98}). Our OCDM predictions seem to be systematically
a little higher than the N-body results, but it should be kept in
mind that measurements at different scales are correlated and that
a measurement-bias due to a division of estimators might be present
(\citenp{hg99}).

The accuracy of our method is actually somewhat surprising because
of the inherent approximate nature of the prescription for $Q_3$ (eq.
[\ref{Q3}]) and of the scaling ansatz for
the power spectrum evolution.
Moreover, the weakly nonlinear fluctuations, which do not obey the hierarchical
form with a configuration independent $Q_3$ as in eq. (\ref{Q3def}), must
contribute at some level to the relevant integrals for $S_3$.
To check this, we perform an alternative integration for $K_2$ in eq. 
(\ref{S3cal}) by including only 'nonlinear' modes: taking the lower
limit of integration to be $\ell_{\rm nl}$ instead of $0$, where
$\ell_{\rm nl}$ satisfies 
$4 \pi (\ell_{\rm nl}/r(\chi))^3 P(\ell_{\rm nl}/r(\chi)) = 1$.
Let us call the resulting integral $K_2^{\rm nl}$, and
define $\Delta_{K_2} \equiv |K_2^{\rm nl} - K_2|/K_2$.
We find that $\Delta_{K_2}$ is very similar for all 3 models above,
and is about $10 - 30 \%$ at $\theta_R \sim 1' - 
5'$, reaching about $45 \%$ at $\theta_R \sim 10'$. 
We therefore propose the following self-consistency check: 
$\Delta_{K_2}$ should be less than about $30 \%$ for our method
to yield reliable estimates of $S_3$.

We show in Fig. \ref{skew_comp}b our prediction of $S_3$ for the
same three models, but the points with error-bars now represent
measurements from simulated surveys, of a size $3^o \times 3^o$, with
$2 \times 10^5$ $z = 1$ galaxies per square degree whose intrinsic
ellipticities are Gaussian distributed with an rms of 0.4 for each
component (attainable with multiple several-hour-long exposures on a
4-meter class telescope using large CCDs; \citenp{ludo99}, 
JSW). It is clear that such a survey can separate these 3 models very nicely. 
Note, however, systematic errors have not been taken into account.

Perhaps more interestingly, we show in the same figure our prediction
of $S_3$ for a cluster-normalized (\citenp{wang98}) quintessence model
(qCDM, see Table \ref{table}).
The equation of state ($w_q = -0.5$) is motivated by certain models of
dynamical supersymmetry breaking (e.g. \citenp{binetruy98}).
Fluctuations in the quintessence component have been ignored, which
is probably a good approximation on the small scales of interest
(\citenp{tw97,caldwell98}). 
Wang et al. \cite*{wang99} have argued that current observations
cannot tell apart qCDM models with $-1 < w_q \,
\approxlt \, -0.4$ from LCDM models, for $\Omega_m \sim 0.3$.
Future microwave background experiments 
would be able to provide better constraints, but there
exist significant degeneracies, especially if $H_0$ is allowed
to vary (\citenp{huey98}). The skewness has the advantage that
it is independent of $H_0$.
Fig. \ref{skew_comp}b shows that the small
angular-scale convergence skewness (especially at $\theta_R \sim 1' -
5'$) provides a promising way to disentangle qCDM and LCDM models: 
fixing $\Omega_m$, $S_3$ varies smoothly from the qCDM values shown
to the LCDM values as $w_q$ changes from $-0.5$ to $-1$.
Moreover, it is a very clean test, because the small-scale $S_3$ is
almost independent 
of all cosmological parameters except $\Omega$'s and $w_q$.

To emphasize this point, we show in Fig. \ref{skewQ} $S_3$'s for 3
different qCDM models, with different $\Gamma$ or $\sigma_8$. They all agree
to within a few percent. From eq.
(\ref{S3cal}) and (\ref{Q3def}), it is not hard to see that the
normalization of $P$ gets divided out in the combination for $S_3$. However,
this really refers to the normalization of the nonlinear $P$. 
The normalization of the linear $P$, $\sigma_8$, should have some
effect on $S_3$ through its impact on the shape of the
nonlinear power spectrum. However, we find that 
at the small scales which dominate the relevant integrals for $S_3$,
the nonlinear power spectra for most models 
have rather similar shapes: a slope around $-1.5$ or so.
Together with the fact that the nonlinear $Q_3$ is a weak function of
scale, this explains why the small angular-scale $S_3$ is relatively
insensitive to both $\sigma_8$ and $\Gamma$. By the same reasoning,
$S_3$ is quite independent of the spectral tilt as well. Nonlinear clustering
seems to erase memory of the initial conditions in $S_3$, 
as far as CDM-type power spectra are concerned. We therefore have in hand
a powerful statistical measure: the small-scale $S_3$ is almost purely
determined by the cosmological energy contents.

\section{Discussion}
\label{discuss}

\begin{figure}[htb]
\centerline{\psfig{figure=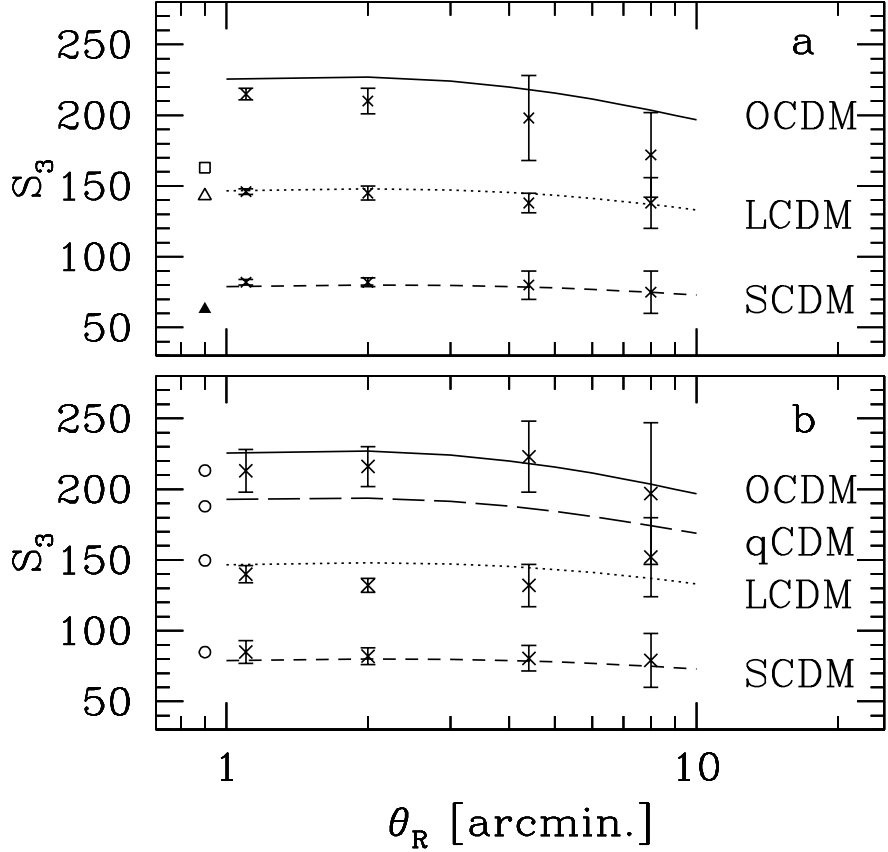,height=5.0in}}
\caption{Panel a: A comparison of the skewness obtained from N-body simulations
(points with error-bars, from 5 - 10 ray-tracing realizations with no
noise added; taken from JSW) and from numerical integration of eq.
(\ref{S3cal}) (lines). The points without error-bars on the far left
denote, from top to bottom, the
perturbation theory predictions of $S_3$ for O/L/SCDM models at
$\theta_R = 1'$ (from JSW). 
Panel b: The lines are the same as before,  
with the addition of predictions for a qCDM model. Points with
error-bars denote measurements for 
O/L/SCDM models from simulated $3^o \times 3^o$ surveys with $2 \times
10^5$ galaxies per sq. deg. (and random noise added; taken from JSW).
The open circles on the far left denote the approximate $S_3$ from eq.
(\protect{\ref{S3approx}}). 
All sources/galaxies are assumed to be at $z = 1$. }
\label{skew_comp}
\end{figure}

\begin{figure}[htb]
\centerline{\psfig{figure=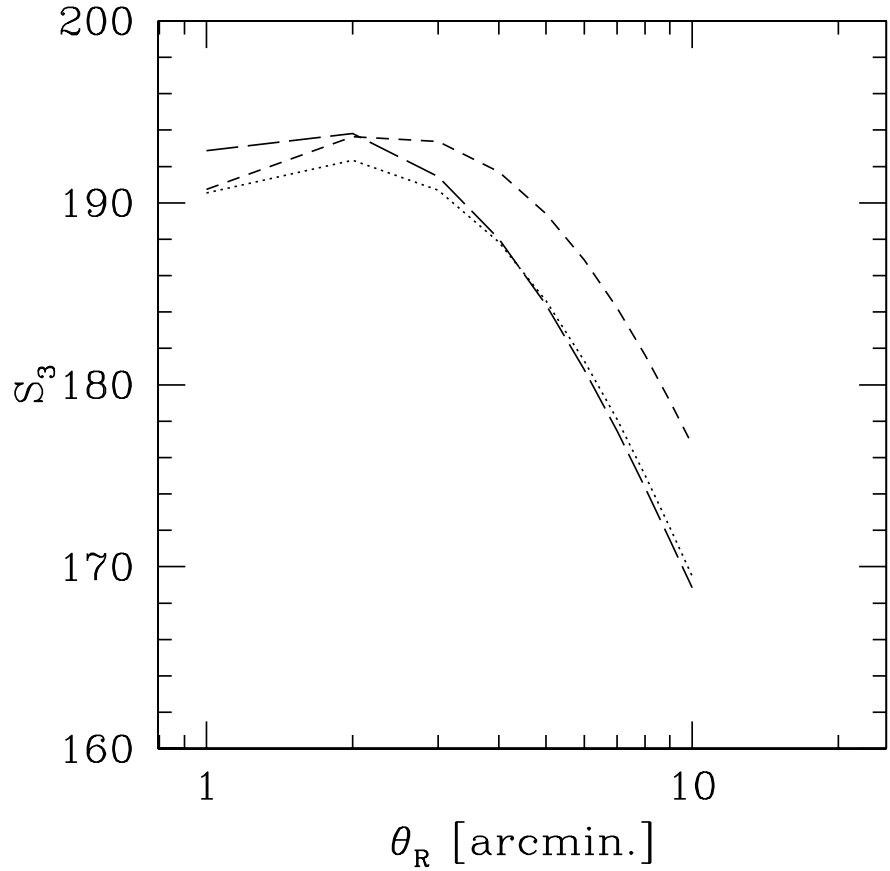,height=5.0in}}
\caption{Skewness for three qCDM models. Long-dashed line - fiducial
qCDM as in Fig. \ref{skew_comp}b (Table \ref{table}); dotted line -
same qCDM but with $\Gamma = 0.5$; short-dashed line - same qCDM but
with $\sigma_8 = 1.0$} 
\label{skewQ}
\end{figure}

How do we understand the cosmological dependence of $S_3$?
The best way is to go back to the definition of the projected mass 
density $\kappa$ in eq. (\ref{kappadef}).
Observe that $w$ occurs three times in the numerator of $S_3$ and
four times in the denominator (eq. [\ref{S3cal}]). This means any
overall constant multiplying $w$ is going to show up in $S_3$
(except for $H_0 /c$ which is canceled out in the combination
for $\kappa$). Among other things, $S_3$ scales as $1/\Omega_m$, the
reciprocal of the total matter 
content. This means flat matter dominated models generally have
lower $S_3$ compared to low density models, as is seen in Fig.
\ref{skew_comp}. Moreover, a $\Lambda$ or quintessence
dominated universe has a larger volume out to $z = 1$, compared
to an open universe, making $w$ larger and $S_3$ smaller. Finally, the
different 
fluctuation growth rates in different cosmologies also
shift the skewness to some extent.
We find that the following crude approximation works 
surprisingly well in reproducing our results from integrating eq.
(\ref{S3cal}):
\begin{eqnarray}
\label{S3approx}
S_3 \sim && 3 \tilde
Q_3 \int_0^{\chi_s} d\chi 
[w(\chi)^3 /r(\chi)^4] [g(z)/(1+z)]^4 r(\chi)^{-2 \tilde n} /
\\ \nonumber
&& \left[ \int_0^{\chi_s} d\chi [w(\chi)^2 / r(\chi)^2]
[g(z)/(1+z)]^2 r(\chi)^{-\tilde n} \right]^2
\end{eqnarray}
where $g(z)$ is given by eq. (\ref{gdef}) or its generalization to
include quintessence, $\tilde Q_3$ is taken to be $2.7$, and
$\tilde n$ is $-1.2$ i.e. the power spectrum is
assumed to obey a power-law with simply linear evolution.
The assumptions of a constant $Q_3$ and a power-law $P$ implies
a scale-independent $S_3$.
The results of applying eq. (\ref{S3approx}) are shown as open circles in
Fig. \ref{skew_comp}b. 

For more accurate results, we recommend going back to eq.
(\ref{S3cal}), (\ref{Q3}) and (\ref{K3approx}), which give $S_3$
accurate to within $10 \%$ 
at $\theta_R \sim 1' - 5'$, assuming the sources are at $z = 1$. 
We have also suggested in \S \ref{skewness} a useful consistency
check: $\Delta_{K_2}$, a measure that quantifies the degree of
linearity, should be less than about $30 \%$. For models
that are too linear, the hierarchical ansatz with a configuration
independent $Q_3$ (eq. \ref{Q3}) breaks down.
An interesting example is provided by the $\tau$CDM model simulated
by JSW, which is exactly the same as SCDM except that 
$\Gamma = 0.21$ and hence has less power on small scales.
As explained before, the highly nonlinear $S_3$ should
be insensitive to $\Gamma$, and our method would predict 
a $\tau$CDM $S_3$ very close to that of SCDM. JSW found that
that the $\tau$CDM N-body results are in fact about $30 \%$ higher
than the SCDM results at a few arcminutes.
Applying our consistency test shows that $\Delta_{K_2}$ is 2 - 3 times
higher for $\tau$CDM compared to all other models we have considered.
Hence, the somewhat large difference between $\tau$CDM and SCDM
seen in JSW is a reflection of their different levels of
nonlinearity. Models with as little small-scale power as
$\tau$CDM are probably inconsistent with 
observations of the Lyman-alpha forest (\citenp{hui97,croft98}).

We have argued that deep lensing surveys, with a total area of several
square degrees and background galaxies at $z \sim 1$, should be
capable of distinguishing between cosmological models with different
energy contents. In particular, contrary to what is indicated
by perturbation theory, the small-scale skewness can be used
to differentiate between curvature and cosmological constant dominated
models. Moreover, $S_3$ also shows sensitivity to the equation of state,
$w_q$, of quintessence models as well, making them distinguishable
from $\Lambda$ models.
In practice, however, since we have only one observable in $S_3$,
one needs to impose extra constraints 
to restrict the range of models when engaging in model testing.
For example, one can assume the class of flat tracker-field models
(\citenp{zlatev98}) where $w_q$ is determined by $\Omega_q$, and
so the only free parameter is $\Omega_q$. Another example: making
use of the fact that for a given $\Omega_m$ curvature dominated models yield
higher $S_3$ than $\Lambda$ models, one can
obtain lower limits on $\Omega_m$. Additional
constraints from other observations such as
the microwave background and large scale structure are obviously
useful.
It should also be borne in mind that systematic errors, such
as those due to the correction of an anisotropic point-spread-function, have
not been taken into account (\citenp{kaiser95}). In addition,
we have assumed that the galaxy redshifts are known, but this is likely
achievable by photometric techniques.

A few issues are worth further investigation. A fitting formula for
the three-point function, which smoothly interpolates
between the perturbative and the highly nonlinear regimes,
could in principle be used to extend our calculation to cover all
angular scales. At present, no such formula exist for 
CDM-type spectra (\citenp{sf98}). Moreover, our method can be easily
generalized to $S_N$ for arbitrary $N$. Such a calculation would
be useful for the estimation of measurement errors from lensing
surveys (\citenp{szh99}). Lastly, our expressions are easily generalizable to
measures such as the aperture mass (\citenp{schneider98}), which
corresponds to using a different smoothing kernel $W_{\theta_R}$ in
eq. (\ref{S3def}), and its Fourier transform $\tilde W$ in the rest of
our expressions. Eq. (\ref{S3approx}) should remain roughly valid
because it is independent of $\tilde W$. 

The author is indebted to Rom\'an Scoccimarro and Matias Zaldarriaga
for helpful comments and for an earlier collaboration which
motivated the present work. Special thanks are due to Rom\'an
Scoccimarro for his very generous help in the course of the
investigation. The author also thanks Enrique
Gazta\~naga for discussions on issues of projection, 
Martin White for discussions on interpretations of N-body simulations,
and Zoltan Haiman and Scott Dodelson for useful comments.
This work was supported by the DOE and the NASA
grant NAG 5-7092 at 
Fermilab.


\vspace{3.0in}

\begin{table}[htb]
\begin{center}
\begin{tabular}{|ccccccc|}\hline
Model & $\Omega_m$ & $\Omega_k$ & $\Omega_q$/$\Omega_\Lambda$ & $w_q$
& $\Gamma$ & $\sigma_8$ \\ \hline \hline
SCDM & 1 & 0 & 0 & -- & 0.5 & 0.6 \\
OCDM & 0.3 & 0.7 & 0 & -- & 0.21 & 0.85 \\
LCDM & 0.3 & 0 & 0.7 & -1 & 0.21 & 0.9 \\
qCDM & 0.3 & 0 & 0.7 & -0.5 & 0.21 & 0.8 \\ \hline
\end{tabular}
\end{center}
\caption{\label{table} A list of models.}
\end{table}

\newpage

\end{document}